\documentclass{article}
\usepackage{amsmath,epsfig,amssymb}
\usepackage{graphicx,color}
\usepackage{eufrak,mathrsfs}
\usepackage[mac]{inputenc}
\usepackage{array} 
\usepackage{upgreek,bm}

\def\0{\boldsymbol{0}}
\def\1{\boldsymbol{1}}
\def\x{\bm{x}}
\def\y{\bm{y}}

\def\0{\boldsymbol{0}}
\def\1{\boldsymbol{1}}
\def\bw{\boldsymbol{\omega}}
\def\u{\boldsymbol u}

\def\k{{\boldsymbol k}}

\def\d{{\boldsymbol d}}

\def\d{\mathrm d}
\def\htau{\acute{\tau}}
\def\D{\mathrm D}

\def\I{\mathscr I}

\def\H{\mathpzc H}
\def\Hil{\mathscr H}

\newcommand{\sinc}{{\rm sinc}}
\newcommand{\mbb}{\mathbb}

\newcommand{\w}{\omega}

\newcommand{\btau}{\boldsymbol{\tau}}

\DeclareSymbolFont{EUr}{U}{eur}{m}{n}
\DeclareSymbolFont{EUb}{U}{eur}{b}{n}
\DeclareMathSymbol{\varphi}{\mathord}{EUr}{"27}
\DeclareMathAlphabet{\mathpzc}{OT1}{pzc}{m}{it}
\DeclareMathOperator{\sign}{\mathrm{sign}} 

\newtheorem{theorem}{Theorem}[section]

\newtheorem{proposition}[theorem]{Proposition}

\begin{document}

\title{On the Shiftability of Dual-Tree Complex Wavelet Transforms}

\author{Kunal Narayan Chaudhury and Michael Unser \thanks{Corresponding Author: Kunal~Narayan~Chaudhury. The authors are with the Biomedical Imaging Group, \'Ecole polytechnique fédérale de Lausanne (EPFL), Station-17, CH-1015 Lausanne VD,
Switzerland. Fax: +41 21 693 37 01, e-mail: \{kunal.chaudhury, michael.unser\}@epfl.ch. This work was supported in part by the Swiss National Science Foundation under grant 200020-109415.}
}

\date{}
\maketitle

\begin{abstract}
		
		The dual-tree complex wavelet transform (DT-$\mbb{C}$WT) is known to exhibit better shift-invariance than the conventional discrete wavelet transform. We propose an amplitude-phase representation of the DT-$\mbb{C}$WT which, among other things, offers a direct explanation for the improvement in the shift-invariance. The representation is based on the \textit{shifting} action of the group of fractional Hilbert transform (fHT) operators, which extends the notion of arbitrary phase-shifts from sinusoids to finite-energy signals (wavelets in particular). In particular, we characterize the \textit{shiftability} of the DT-$\mbb{C}$WT in terms of the shifting property of the fHTs. At the heart of the representation are certain fundamental invariances of the fHT group, namely that of translation, dilation, and norm, which play a decisive role in establishing the key properties of the transform. It turns out that these fundamental invariances are exclusive to this group.  
	 	 
	Next, by introducing a generalization of the Bedrosian theorem for the fHT operator, we derive an explicitly understanding of the shifting action of the fHT for the particular family of wavelets obtained through the modulation of lowpass functions (e.g., the Shannon and Gabor wavelet). This, in effect, links the corresponding dual-tree transform with the framework of windowed-Fourier analysis.  
	
	Finally, we extend these ideas to the multi-dimensional setting by introducing a directional extension of the fHT, the fractional directional Hilbert transform. In particular, we derive a signal representation involving the superposition of direction-selective wavelets with appropriate phase-shifts, which helps explain the improved shift-invariance of the transform  along certain preferential directions.		 
				 
\end{abstract}

\section{INTRODUCTION}
\label{intro}

	The dual-tree complex wavelet transform (DT-$\mbb{C}$WT) is an enhancement of the conventional discrete wavelet transform (DWT) that has gained increasing popularity as a signal processing tool. The transform, originally proposed by Kingsbury \cite{kingsbury2} to circumvent the shift-variance problem of the decimated DWT, involves two parallel DWT channels with the corresponding wavelets forming approximate Hilbert transform pairs \cite{CTDWT}. We refer the reader to the excellent tutorial \cite{CTDWT} on the design and application of the DT-$\mbb{C}$WT.

	In this contribution, we characterize the dual-tree transform from a complementary perspective by formally linking the multiresolution framework of wavelets with the amplitude-phase representation of Fourier analysis. The latter provides an efficient way of encoding the relative location of information in signals through the phase function that has a straightforward interpretation. Specifically, consider the Fourier expansion of a finite-energy signal $f(x)$ on $[0,L]$:
\begin{align}
\label{fourier}
f(x)= a_0&+a_1 \cos(\w_0 x)+a_2 \cos(2\w_0 x)+ \cdots \nonumber \\
&+ b_1 \sin(\w_0 x)+b_2 \sin(2\w_0 x)+ \cdots.
\end{align}
Here $\w_0 \ (\w_0L=2\pi)$ denotes the fundamental frequency, and $a_0, a_1,a_2,\ldots$, and $b_1,b_2,\ldots$ are the (real) Fourier coefficients corresponding to the  even and odd harmonicsrespectively. Now, by introducing the complex Fourier coefficients $c_n=a_n-jb_n$ and by expressing them in the polar form $c_n= |c_n| e^{j \phi_n}, 0 \leqslant  \phi_n < 2\pi$, one can rewrite \eqref{fourier} as
\begin{align}
\label{f2}
f(x)&= \sum_{n=0}^{\infty} |c_n| \Big(\cos \phi_n \cos(n\w_0x)-\sin \phi_n \sin(n\w_0x)\Big) = \sum_{n=0}^{\infty}  |c_n| \varphi_n\big(x+\htau_n\big)
\end{align}
with $\htau_n=\phi_n/n\w_0$ specifying the displacement of the reference sinusoid $\varphi_n(x)=\cos(n\w_0x)$ relative to its fundamental period $[0,L/n]$. The above amplitude-phase  representation highlights a fundamental attribute of the shift parameter $\htau_n$: it corresponds to the shift $\htau$ that maximizes $|\langle f(\cdot), \varphi_n(\cdot+\htau) \rangle|$, the correlation of the signal with the reference $\varphi_n(x)$. The corresponding amplitude $|c_n|$  measures the strength of the correlation.

	As far as signals with isolated singularities (e.g. piecewise-smooth signals) are concerned, the wavelet representation -- employing the dilated-translated copies of a fast-decaying oscillating waveform -- has proven to be more efficient. Moreover, the added aspect of multiscale representation allows one to zoom onto signal features at different spatial resolutions. Complex wavelets, derived via the combination of non-redundant wavelet bases, provide an attractive means of recovering the crucial phase information. In particular, the phase relation between the components (of the complex wavelet) is used to encode the relative signal displacement (besides offering robustness to interference). The DT-$\mbb{C}$WT is a particular instance where the components are related via the Hilbert transform \cite{CTDWT,selesnick}.

\subsection{Main Results}

	Analogous to the fact that the complex Fourier coefficients in \eqref{f2} are derived from the (primitive) analytic signals $\mathrm{e}^{jn\w_0x}=\varphi_n(x)+j\Hil \varphi_n(x)$:
\begin{equation*}
c_n=a_n-jb_n=\langle f(x), \mathrm{e}^{jn\w_0x} \rangle_{[0,L]},
\end{equation*}
the DT-$\mbb{C}$WT coefficients are obtained by projection the signal on to the dilated-translated copies of the analytic wavelet $\Psi(x)=\psi(x)+j\Hil \psi(x)$. The central idea of this paper is the identification of the wavelet counterparts of the phase-shifted sinusoids $\varphi_n(x+\tau_n)$ in \eqref{f2}. In particular, these so-called \textit{shifted} wavelets  are derived by the action of the single-parameter family of fHT operators $\{\H_{\tau}\}_{\tau \in \mathbf{R}}$ (for definition and properties see \S\ref{defs}) on the reference wavelet $\psi(x)$. The shift parameter $\tau$ controls the \textit{shifting} action $\psi(x) \mapsto \H_{\tau} \psi(x)$ of the fHT, and, in effect, resulting in the realization of a continuously-defined family of shifted wavelets. This action has an important connotation in relation to pure sinusoids: $\H_{\tau}\varphi_n(x)=\cos(n\w_0x+\pi\tau)$. In fact, the amplitude-phase representation of the DT-$\mbb{C}$WT is derived in \S\ref{shiftability} by generalizing the following equivalent expression of \eqref{f2}: 
\begin{equation*}
f(x)=\sum_{n=0}^{\infty} |c_n| \ \H_{\tau_n} \varphi_n(x) \quad \quad (\tau_n=\phi_n/\pi)
\end{equation*}
which, in turn, is based on the above-mentioned phase-shift action of the fHTs. The significance of either representation is that they allow us to provide a precise characterization of the shiftability of the associated reference functions in terms of the shifting action of the fHT. 

	Motivated by this connection, we make a detailed study of the group of fHT operators in \S\ref{properties}. In particular,  we highlight their invariance to translations and dilations which allows us to seamlessly integrate them into the multiresolution framework of wavelets. Moreover, the observation that the above-mentioned invariances are exclusively enjoyed by the fHT group (cf. Theorem \ref{representation}) makes the shiftability of the dual-tree transform unique.  For the particular family of dual-tree transforms involving HT pairs of modulated wavelets, we derive an explicit characterization of the shifting action of the fHT in \S\ref{mod_wavelets}. If the dual-tree wavelet is not modulated, we can still characterize the  action of the fHT by studying the family of fractionally-shifted wavelets $\{\H_{\tau}\psi\}_{\tau \in \mathbf{R}}$, and we do this explicitly for the particular case of spline wavelets in \S\ref{spline}. Finally, we extend the proposed representation to the multi-dimensional setting in \S\ref{directional_wavelets} by introducing certain directional extensions of the fHT. 
	
	The above results have certain practical implications. In \S\ref{metrics}, we propose certain measures for accessing the quality of the factors, namely the HT correspondence and modulation criterion, that are fundamental to the shiftability property of the dual-tree wavelets. These metrics could prove useful in the design of dual-tree wavelets with a good shift-invariance property.

\section{THE FRACTIONAL HIBERT TRANSFORM}
\label{defs}

	In  what follows, the Fourier transform of a function $f(\x)$ defined over $\mathbf{R}^d \ (d \geqslant 1)$ is specified by $\hat{f}(\bw)=\int_{\mathbf{R}^d} f(\x) \exp{(-j \bw^T \x)} d\x$, where $\bw^T \x$ denotes the usual inner-product on $\mathbf{R}^d$. The other transform that plays a significant role is the Hilbert transform (HT) \cite{Bracewel,Stein_Weiss}; we will denote it by $\Hil$. In particular, we shall frequently invoke the Fourier equivalence
\begin{equation}
\label{HT_def}
\Hil f(x) \stackrel{\mathscr{F}}{\longleftrightarrow} -j \sign(\w)\hat{f}(\w)
\end{equation}
characterizing the action of the HT on $\mathrm{L}^2(\mathbf{R})$, the class of finite-energy signals\footnote{The domain can also be extended to include distributions such as the Dirac delta and the sinusoid \cite[Chapter 2]{harmonic_analysis}}. Three fundamental properties of the HT that  follow from \eqref{HT_def} are its invariance to translations and dilations, and its unitary (norm-preserving) nature. Moreover, we shall use $\I$ to denote the identity operator $(\I f)(x)=f(x)$.

	We begin with a detailed exposition of the relevant characteristics of the fHT that forms the cornerstone of the subsequent discussion. There exit several definitions of the fHT in the signal processing and optics literature \cite{SPLetters_fHT2,SPLetters_fHT1,optics_fHT,Davis_fHT}; however, for reasons that will be obvious in the sequel, we propose to formulate it as an interpolation of the ``quadrature'' identity and HT operator using conjugate trigonometric functions. In particular, we define the fHT operator $\H_{\tau},$ corresponding to the real-valued \textit{shift} parameter $\tau$, as 
\begin{equation}
\label{def_fHT}
\H_{\tau}=\cos(\pi\tau)\ \I - \sin(\pi\tau) \ \Hil.
\end{equation}

	This definition is equivalent to the formulation introduced in \cite{optics_fHT,Davis_fHT}, but differs from the ones in \cite{SPLetters_fHT2,SPLetters_fHT1} up to a complex chirp. The important aspect of the above operator-based formulation is that it directly relates the fHT and its properties to the more fundamental identity and HT operator, which are identified \textit{a posteriori} as special instances of the fHT: $\I=\H_0$ and $\Hil=\H_{-1/2}$. In view of \eqref{def_fHT}, we would like to make note of fact that $\I$ and $\H$ have a non-linear correspondence:
\begin{equation}
\label{relation_id_HT}
\Hil^2=-\I,
\end{equation}
and act in ``quadrature'' in the sense that	
\begin{equation}
\label{quadrature_corresp}
\langle \I f, \Hil f\rangle=0 \qquad (f \in \mathrm{L}^2(\mathbf{R})).
\end{equation}
These come as a direct consequence of definition \eqref{HT_def} and certain properties of the inner-product.
			
	As far as the domain of definition of \eqref{def_fHT} is concerned, note that both $\I$ and $\Hil$ act as bounded operators (with a bounded inverse) on $\mathrm{L}^p(\mathbf{R})$ for $1 <p< \infty$  \cite{harmonic_analysis}, and so does $\H_{\tau}$. In particular, the fHT admits the following equivalent specification on $\mathrm{L}^2(\mathbf{R})$: 
\begin{equation}
\label{freq_response}
\H_{\tau}f(x) \stackrel{\mathscr{F}}{\longleftrightarrow}  \exp\big(j\pi \tau \sign(\w)\big)\hat f(\w)
\end{equation}
that comes as a consequence of equivalence \eqref{HT_def}. We shall henceforth invoke \eqref{def_fHT} and \eqref{freq_response} interchangeably in the context of finite-energy signals.

\subsection{Characterization of the fHT}	
\label{properties}

 	As remarked earlier, most of the characteristic features of the constituent identity and HT operators carry over to the family of fHT operators $\{\H_{\tau}\}_{\tau \in \mathbf{R}}$. In particular, the following properties of the fHT can be readily derived:\newline 
$\mathrm{(P1)}$ Translation-invariance:  $(\H_{\tau} f)(x-y)=\H_{\tau}\{f(\cdot-y)\}(x)$ for all real $y$.\\
$\mathrm{(P2)}$ Dilation-invariance: $(\H_{\tau} f)( \lambda x)=\H_{\tau}\{f(\lambda \cdot)\}(x)$ for all positive $\lambda$.\\
$\mathrm{(P3)}$ Unitary nature: $\langle\H_{\tau} f,\H_{\tau} g\rangle=\langle f, g\rangle$; in particular, $||\H_{\tau}f||=||f||$ for all $f,g$ in $\mathrm{L}^2(\mathbf{R})$.\\
$\mathrm{(P4)}$ Composition law: $\H_{\tau_1}\H_{\tau_2}=\H_{\tau_1+\tau_2}$.\\
$\mathrm{(P5)}$ Phase-shift operator: $\H_{\tau} \{\cos(\w_0x)\}=\cos(\w_0x+\pi \tau)$.\newline

Indeed, (P$1$) and (P$2$) are immediate consequences of the dilation- and translation-invariance of $\I$ and $\Hil$; (P$3$) follows from the unitary frequency response  \eqref{freq_response} and Parseval's identity; and (P$4$) follows from \eqref{relation_id_HT}:
\begin{align*}
\H_{\tau_1} \H_{\tau_2} &=\big(\cos(\pi\tau_1)\ \I - \sin(\pi\tau_1) \ \Hil\big) \big(\cos(\pi\tau_2)\ \I - \sin(\pi\tau_2) \ \Hil\big) \nonumber \\
&=\cos\big(\pi(\tau_1+\tau_2)\big)\ \I - \sin\Big(\pi(\tau_1+\tau_2)\big) \ \Hil.
\end{align*}
Finally, it is the quadrature-shift action $\cos(\w_0x) \mapsto \sin(\w_0x)$ of $\Hil$ that results in (P$5$): 
\begin{align}
\label{phase-shift}
\H_{\tau} \cos(\w_0x)&=\cos(\pi \tau) \cos(\w_0x)-\sin(\pi \tau) \Hil \{\cos(\w_0x)\}=\cos(\w_0x+\pi \tau);
\end{align}
this is also justified by the frequency response of the fHT.

\begin{figure}
\centering
\begin{tabular}{cc}
\fbox{\includegraphics[width=0.3\linewidth]{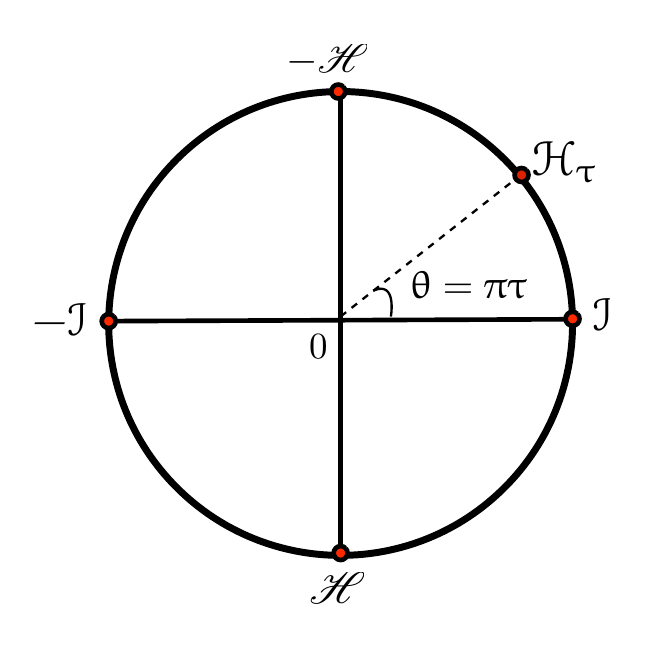}}
\end{tabular}
\caption{Geometrical interpretation of the continuous fHT group using the isomorphic unit-circle group $S^1$ (the multiplicative group of complex numbers having unit modulus); the correspondence is $\H_{\tau} \longleftrightarrow (\cos(\pi \tau),\sin(\pi \tau)) $.}
\label{fig1}
\end{figure}

		As will be discussed shortly, properties (P$1$), (P$2$) and (P$3$) play a crucial role in connection with wavelets. The composition law (P$4$) tells us that the family of fHT operators is closed with respect to composition. Moreover, as the identity $\H_{\tau}\H_{-\tau}=\H_0=\I$ suggests, the inverse\footnote{Note that, for a given $\tau$, there exists infinitely many $\tau'$ such that identity $\H_{\tau}\H_{\tau'}=\I$ holds; this comes as a consequence of the periodicity of the trigonometric functions involved in the definition. One can easily factor out the periodic structure by identifying $\tau_1$ and $\tau_2$ iff $\H_{\tau_1}=\H_{\tau_2}$;  in effect, this equivalence relation results in the specification of equivalence classes of fHTs. However, for the simplicity of notation, we shall henceforth use $\H_{\tau}$ to denote both the equivalence class and its representatives.} fHT operator is also a fHT operator specified by $\H^{-1}_{\tau}=\H_{-\tau}$. These closure properties can be summarized by following geometric characterization: 
\vspace{0.05in}		
\begin{proposition} 
The family of fHT operators forms a commutative group on $\mathrm{L}^p(\mathbf{R}), 1 <p< \infty$.  	
\end{proposition}	
\vspace{0.05in}		
	 In this respect, note the marked resemblance between the family of fHT operators and the commutative group of translation operators that play a fundamental role in Fourier analysis; the relevance of the former group in connection with the DT-$\mbb{C}$WT will be demonstrated in the sequel. Moreover, the finite subgroup $\{\I,\H,-\I,-\H\}$ of self-adjoint\footnote{self-adjoint up to a sign: $T^{*}=\pm T$ for each $T$ in the subgroup.} operators (see Fig. \ref{fig1}) is worth identifying. It is the smallest subgroup containing the in-phase/quadrature operators that play a fundamental role in the dual-tree transform.

\subsection{The Wavelet Context}
	
	Similar to the Hilbert transform, the fHT perfectly fits the wavelet framework. The implication of properties (P$1$) and (P$2$) is that the fHT of simultaneous dilates and translates of a wavelet is a wavelet, dilated and translated by the same amount. This has fundamental ramifications in connection with dyadic wavelet bases generated via the dilations and translations of a single mother-wavelet $\psi(x)$. In particular, let $\psi_{i,k}(x)$ and $\Xi_{i,k}$ denote the dilated-translated wavelets $\Xi_{i,k} \psi(x)=2^{i/2} \psi(2^i x-k), (i,k) \in \mathbf{Z}^2$ and the corresponding (normalized) dilation-translation operators, respectively. Then the commutativity
\begin{equation}
\label{inv}
\H_{\tau} \  \Xi_{i,k}=\Xi_{i,k} \H_{\tau}
\end{equation}
holds for all real $\tau$ and integers $i$ and $k$. The significance of \eqref{inv} is that it allows us to conveniently factor out the pervasive dilation-translation structure while analyzing the action of fHTs on wavelet bases.  
	
 	On the other hand, a fundamental consequence of the isometry property (P$3$) is that $\H_{\tau}$ maps a Riesz basis onto a Riesz basis; in particular, if $\{\psi_{i,k}\}$ forms a wavelet basis of $\mathrm{L}^2(\mathbf{R})$, then so does $\{\H_{\tau} \psi_{i,k}\}$. In fact, $\H_{\tau}$ preserves biorthogonality: if $\{\psi_{i,k}\}$ and $\{\tilde\psi_{i',k'}\}$ constitute a biorthogonal wavelet basis satisfying the duality criteria $\langle \psi_{i,k},\tilde \psi_{i',k'} \rangle=\delta[i-i'] \delta[k-k']$, then also we have that
\begin{equation*}
\langle \mathcal{H}_{\tau} \psi_{i,k},\mathcal{H}_{\tau} \tilde\psi_{i',k'} \rangle=\langle \psi_{i,k},\tilde\psi_{i',k'} \rangle=\delta[i-i'] \delta[k-k'],
\end{equation*}
signifying that $\{\mathcal{H}_{\tau}\psi_{i,k}\}$ and $\{\mathcal{H}_{\tau}\tilde\psi_{i',k'}\}$ form a biorthogonal basis as well.

\section{SHIFTABILITY OF THE DUAL-TREE TRANSFORM}

\subsection{Multiscale Amplitude-Phase Representation}
\label{shiftability}

	We now derive the amplitude-phase representation of the DT-$\mathbb{C}$WT based on the shifting action of the fHT. As remarked earlier, the parallel is grounded on the observation that instead of the quadrature sinusoids $\{\cos(n\w_0 x)\}$ and $\{\sin(n\w_0 x)\}$, the DT-$\mbb{C}$WT employs two parallel wavelet bases, $\{\psi_{i,k}\}$ and $\{\psi'_{i,k}\}$, derived via the dilations-translations of the wavelets $\psi(x)$ and $\psi'(x)$ that  form a HT pair: $\psi'(x)=\Hil \psi(x)$. A signal $f(x)$ in $\mathrm{L}^2(\mathbf{R})$ is then simultaneously analyzed in terms of  these wavelet bases  yielding the wavelet expansions
\begin{align}
\label{branches}
f(x)= \begin{cases}  \sum_{(i,k) \in \mathbf{Z}^2} a_i[k] \psi_{i,k}(x).   \\
\sum_{(i,k) \in \mathbf{Z}^2} b_i[k] \psi'_{i,k}(x)  .
\end{cases}
\end{align}
The analysis coefficients $a_i[k]$ and $b_i[k]$ are specified by the projections onto the dual wavelet bases $\{\tilde \psi_{i,k}\}$ and $\{\tilde \psi'_{i,k}\}$: 
\begin{equation*}
a_i[k]=\langle f, \tilde \psi_{i,k} \rangle, \quad  \text{and}  \quad b_i[k]=\langle f, \tilde  \psi'_{i,k} \rangle.
\end{equation*}
The dual wavelet bases are implicitly related to the corresponding primal bases through the duality criteria $\langle \tilde \psi_{i,k}, \psi_{p,n} \rangle=\delta[i-p,k-n]$ and $\langle \tilde\psi'_{i,k}, \psi'_{p,n} \rangle=\delta[i-p,k-n]$. A fundamental consequence of the unitary property of the HT is that these dual bases can be generated through the dilations-translations of two dual wavelets, say $\tilde \psi(x)$ and $\tilde \psi'(x)$, that form a HT pair as well: $\tilde \psi'(x)=\Hil\tilde \psi(x)$ \cite{kunal_journal}. In particular, by introducing the complex wavelet 
\begin{equation*}
\quad \tilde \Psi(x)=\frac{1}{2}\big(\tilde \psi(x)+j\tilde \psi'(x)\big)
\end{equation*}
and its dilated-translated versions $\tilde \Psi_{i,k}(x)$ -- the analytic counterpart of the complex sinusoids $\exp(jn\w_0x)$ -- the dual-tree analysis can simply be viewed as the sequence of transformations $f(x) \mapsto c_i[k]=\langle f, \tilde \Psi_{i,k} \rangle$ resulting in the complex analysis coefficients $\{c_i[k]\}_{(i,k) \in \mathbf{Z}^2}$.
 
	 Our objective is to derive a representation of $f(x)$ in terms of the modulus-phase information $c_i[k]=|c_i[k]|\mathrm{e}^{j\phi_i[k]}$, and the reference wavelets $\psi_{i,k}(x)$. In particular, by combining the expansions in \eqref{branches} and by invoking \eqref{inv}, we arrive at the following representation:
\begin{align}
\label{amp-phase}
f(x)&= \frac{1}{2} \sum_{(i,k) \in \mathbf{Z}^2} \Big(a_i[k] \psi_{i,k}(x)+b_i[k] \psi'_{i,k}(x)\Big)   \nonumber \\
& =  \sum_{(i,k) \in \mathbf{Z}^2} |c_i[k]| \ \H_{\phi_i[k]/\pi} \big\{\psi_{i,k}(x)\big\}  \nonumber \\
& = \sum_{(i,k) \in \mathbf{Z}^2} |c_i[k]| \ \Xi_{i,k }\big\{\psi\big(x;\tau_i[k]\big) \big\},
\end{align}
where the synthesis wavelet $\psi\big(x;\tau_i[k]\big)=\H_{\tau_i[k]} \psi(x)$ is derived from the mother wavelet $\psi(x)$ through the action of the fHT corresponding to the shift $\tau_i[k]=\phi_i[k]/\pi$. The above multiresolution amplitude-phase representation provides two important insights into the signal transformation 
\begin{equation*}
f(x) \mapsto \Big\{\big(|c_i[k]|,\tau_i[k]\big)\Big\}_{(i,k) \in \mathbf{Z}^2}.
\end{equation*}
The first of these is derived from the observation that the fractionally-shifted wavelets $\psi(x;\tau_i[k])$ in \eqref{amp-phase} play a role analogous to the phase-shifted sinusoids $\varphi_n(x+\tau_n)$ in \eqref{f2}. In particular, \eqref{amp-phase} offers a rigorous interpretation of the ampitude-phase factors: while $|c_i[k]|$ indicates the strength of wavelet correlation, the relative signal displacement gets encoded in the shift $\tau_i[k]$ corresponding to the most ``appropriate'' wavelet within the family $\{\H_{\tau} \psi_{i,k}\}_{\tau \in \mathbf{R}}$. 
	
	The ``shiftable-wavelet representation'' in  \eqref{amp-phase} also offers an explanation for the improved shift-invariance of the dual-tree transform which is complementary to the frequency-domain argument given by Kingsbury in \cite{kingsbury2}. It is well-known that the signal representation associated with the (critically-sampled) discrete wavelet transform is not shift-invariant; in particular, the uniform sampling of the translation-parameter of the continuous transform limits the degree of shift-invariance of the transform. The fact that the over-sampled dual-tree transform tends to exhibit better shift-invariance can then be explained in terms of the associated shiftability of the transform. Indeed, the fractional-shifts of the reference wavelets around their discrete translates partially compensates for  the limited freedom of translation.

	It is clear that the invariances of the fHT group were central to the derivation of \eqref{amp-phase}. The following result establishes the fHTs as the \textit{only} complete family of operators that exhibits such characteristics:

\vspace{0.05in}	  
\begin{theorem} \emph{(Uniqueness of the fHT)}
\label{representation}

A unitary linear operator $T$ on $\mathrm{L}^2(\mathbf{R})$ is invariant to translations and dilations if and only if it can be represented as
\begin{equation}
T=\cos \theta \ \I-\sin \theta \ \Hil
\end{equation}
for some unique $\theta$ in $(-\pi,\pi]$.
 \end{theorem}
 \vspace{0.05in}	
The above result (proof provided in Appendix \ref{A1}) signifies that any family of unitary operators that simultaneously commutes with translations and dilations is  isomorphic to the fHT group. This provides significant insight into the representation in \eqref{amp-phase} since it is these fundamental invariances that facilitate the incorporation of the fHT into the wavelet framework.

As discussed next, it turns out that the shifted wavelets can be very explicitly characterized for certain classes of wavelets which provides a deeper insight into the above signal representation. 

\begin{figure}
\centering
\begin{tabular}{cc}
\includegraphics[width=0.5\linewidth]{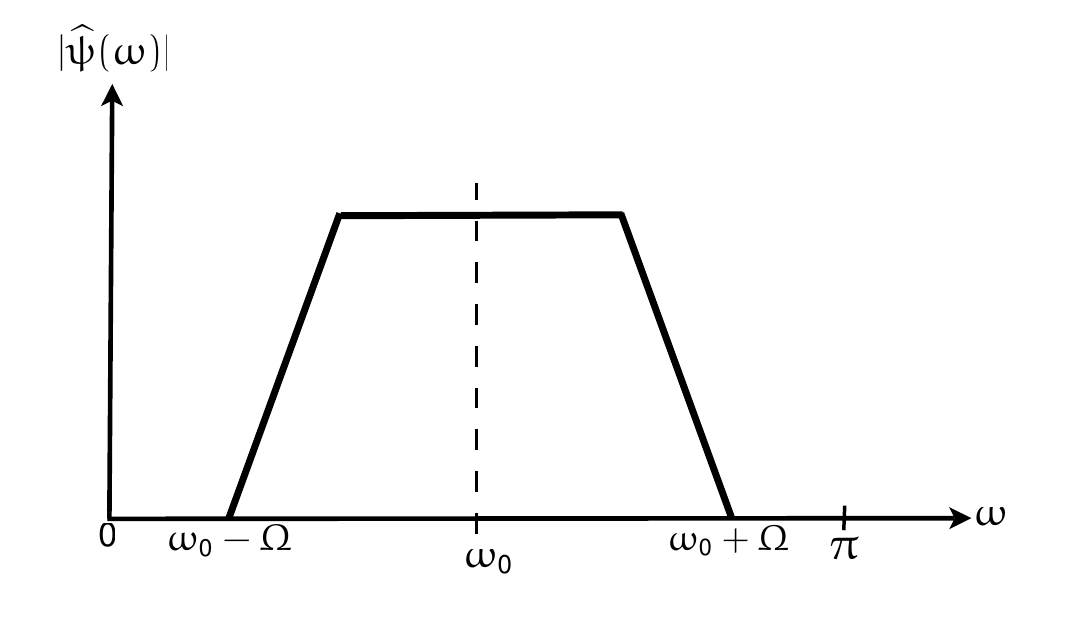}
\end{tabular}
\caption{Idealized spectrum of a modulated wavelet. The spectrum has passbands over $\w_0-\Omega < |\w|< \w_0+\Omega$ with local axes of symmetry at $\w=\pm \w_0$.}
\label{fig2}
\end{figure} 

\subsection{Modulated Wavelets: Windowed-Fourier-like Representation}	
\label{mod_wavelets}

		 A wavelet is, by construction, a bandpass function. Specifically, if the wavelet is a modulated function of the form 
\begin{equation}
\psi(x)=\varphi(x) \cos\big(\w_0 x + \xi_0\big),
\end{equation}
where $\varphi(x)$ is bandlimited to $[-\Omega,\Omega]$ (for some arbitrary $\Omega$), and $\w_0>\Omega$ [cf. Fig. \ref{f2}], we can make precise statements on the representation \eqref{amp-phase} corresponding to the dual-tree transform involving such a modulated wavelet and its HT pair. In order to do so, we provide a very generic result that allows us to extend the phase-action in \eqref{phase-shift} to modulated functions:

\vspace{0.05in}	
\begin{theorem} \emph{(Generalized Bedrosian Identity)}
\label{Bedrosian} 
\vspace{0.05in}	

Let $f(x)$ and $g(x)$ be two real-valued functions such that the support of $\hat{f}(\w)$ is restricted to $(-\Omega,\Omega)$, and that $\hat{g}(\w)$ vanishes for $|\w| < \Omega$ for some arbitrary frequency $\Omega$. Then the fHT of high-pass function completely determines the fHT of the product: 
\begin{equation}
\H_{\tau} \big \{f(x)g(x) \big \}= f(x) \H_{\tau}g(x).
\end{equation}
\end{theorem}

 	Informally, the above result (cf. Appendix \S\ref{AA} for a proof) asserts that the fHT of the product of a lowpass signal and a highpass signal (with non-overlapping spectra) factors into the product of the lowpass signal and the fHT of the highpass signal. Note that, as a particular instance of Theorem \ref{Bedrosian} corresponding to $\tau=-1/2$, we recover the result of Bedrosian \cite{bedrosian} for the HT operator. An important consequence of Theorem \eqref{Bedrosian} is that, for a wavelet is the form $\upvarphi(x) \cos(\w_0x)$ with $\upvarphi(x)$ is bandlimited to $(-\w_0,\w_0)$, the fHT acts on the phase of the modulating sinusoid while preserving the lowpass  envelope:
\begin{equation}
\label{modulation}
\H_{\tau} \big \{ \upvarphi(x) \cos(\w_0x)\big \}=\upvarphi(x) \cos(\w_0x+\pi \tau).
\end{equation}
In particular, the above modulation action allows us to rewrite the signal representation in \eqref{amp-phase} as 
\begin{equation}
\label{WFA}
\sum_{(i,k) \in \mathbf{Z}^2} \stackrel{\mathrm{fixed \ window}}{\overbrace{\upvarphi_{i,k}(x)}}  \Xi_{i,k}  \Big\{ \stackrel{\mathrm{variable \ amp-phase \ oscillation}}{\overbrace{ \big|c_i[k]\big| \cos\big(\w_0 x+\xi_0+\pi \tau_i[k]\big)}}\Big\},
\end{equation}
where $\upvarphi_{i,k}(x)=\Xi_{i,k} \upvarphi(x)$ denotes the (fixed) window at scale $i$ and translation $k$. This provides an explicit interpretation of the parameter $\tau_i[k]$ as the phase-shift applied to the modulating sinusoid of the wavelet. In this regard, its role is therefore similar to that of the shift parameter $\htau_n$ in the Fourier representation \eqref{f2}. In effect,  while the localization window $\upvarphi_{i,k}(x)$ is kept fixed, the oscillation is shifted to best-fit the underlying signal singularities/transitions. In this light, one can interpret the associated dual-tree analysis as a multiresolution form of the windowed-Fourier analysis, with the fundamental difference that, instead of analyzing the signal at different frequencies, it resolves the signal over different scales (or resolutions).

	Two concrete instances of such modulated wavelets are:  
\begin{itemize}
\item \textbf{ Shannon wavelet:} The Shannon wavelet is constructed from the Shannon multiresolution \cite{WTSP}; it is specified as 
\begin{equation*}
\psi(x)=\sinc\left(\frac{x-1/2}{2}\right) \cos\left(\frac{3\pi (x-1/2)}{2}\right),
\end{equation*}
and it dilates-translates constitute an orthonormal wavelet basis of $\mathrm{L}^2{\mathbf{R}}$. Many wavelet families converge to the Shannon multiresolution as the order increases \cite{akram_convergence}; e.g., the orthonormal Battle-Lemarié wavelets \cite{Battle}, and the interpolating  Dubuc-Deslauriers wavelets \cite{DubucWavelets}. 
\item \textbf{Gabor wavelet:} The $\sinc(x)$ envelope of the Shannon wavelet results in an ``ideal'' frequency resolution but only at the expense of poor spatial decay. As against this, wavelets modeled on the Gabor functions \cite{gabor} (modulated Gaussians) exhibit better space-frequency localization. Moreover, as in the case of the Shannon wavelet, they are quite generic in nature since several wavelet families closely resemble the Gabor function. For example, the B-splines wavelets (a semi-orthogonal family of  spline wavelets) asymptotically converge to the Gabor wavelet
\begin{equation*}
\psi(x)=g^{\alpha}(x) \cos\big(\w_0 x+\xi_0\big),
\end{equation*}
where $g^{\alpha}(x)$ is a Gaussian window that is completely determined by the degree of the spline \cite{convergence}. In fact, based on this observation, a multiresolution Gabor-like transform was realized in \cite{kunal_journal} within the framework of the dual-tree transform. This Gabor-like transform involved the computation of the projections
\begin{equation}
f(x) \mapsto \left\langle f(x), \Xi_{i,k} \Psi(x) \right \rangle \qquad (i,k \in \mathbf{Z})
\end{equation}
on to the dilates-translates of the Gabor-like wavelet $\Psi(x)=\psi(x)+j\Hil \psi(x)$, and was realized using the usual dual-tree transform corresponding to the spline wavelets $\psi(x)$ and $\Hil \psi(x)$. We would, however, like to point out that the representation in \eqref{WFA} corresponds to a situation where the role of the analysis and synthesis functions have been reversed, namely one in which the signal is analyzed using the dual complex wavelet $\tilde \Psi(x)$, and where the Gabor-like wavelet $\Psi(x)$ is used for reconstruction. 
\end{itemize}

\begin{figure*}
\centering
\begin{tabular}{cc}
\includegraphics[width=0.8\linewidth]{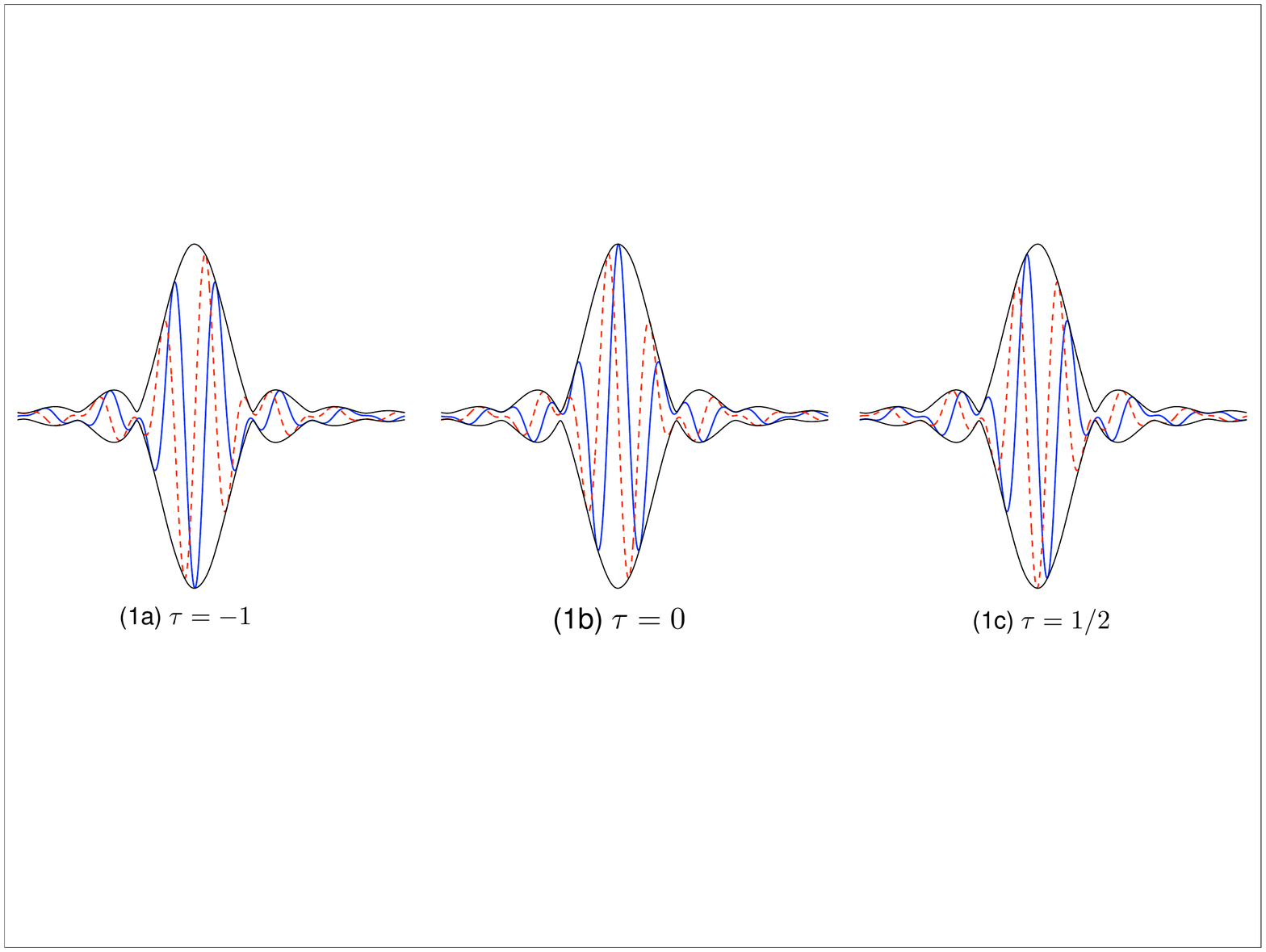} \\
\includegraphics[width=0.8\linewidth]{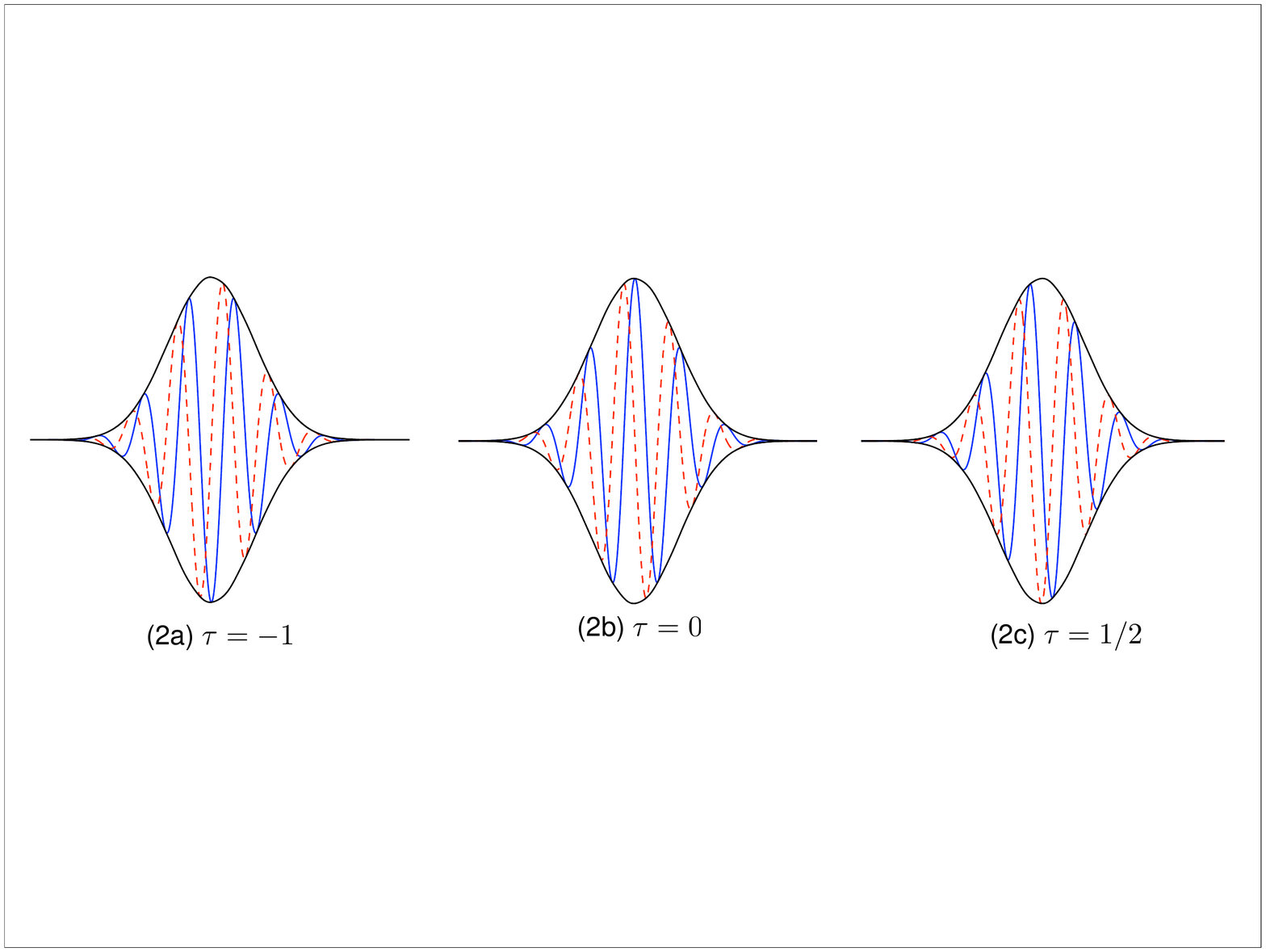} 
\end{tabular}
\caption{Quadrature pairs of orthonormal spline  (resp. B-spline) wavelets resembling the Shannon (resp. Gabor) wavelet: Solid (blue) graph: $\H_{\tau} \psi^8(x)$; Broken (red) graph: $\H_{\tau+1/2} \psi^8(x)$; and Solid (black) graph: Common localization window given by  $|\H_{\tau} \psi^8(x) +j\H_{\tau+1/2}  \psi^{8}(x)|$.}
\label{fig3}
\end{figure*}

\vspace{0.05in}	
\textbf{Graphical Illustrations}: Fig. \ref{fig3} shows quadrature pairs $\left(\H_{\tau} \psi(x),\H_{\tau+1/2}  \psi(x)\right)$ of Shannon-like (resp. Gabor-like) wavelets corresponding to different $\tau$. The fHT pair $(\H_{\tau},\H_{\tau+1/2})$ drives the modulating oscillation to a relative quadrature that are localized within a common $\mathrm{sinc}$-like (resp. Gaussian-like) window specified by $|\H_{\tau} \psi(x)+j\H_{\tau+1/2}  \psi(x)|$.

\begin{figure*}
\centering
\includegraphics[width=100mm,height=100mm]{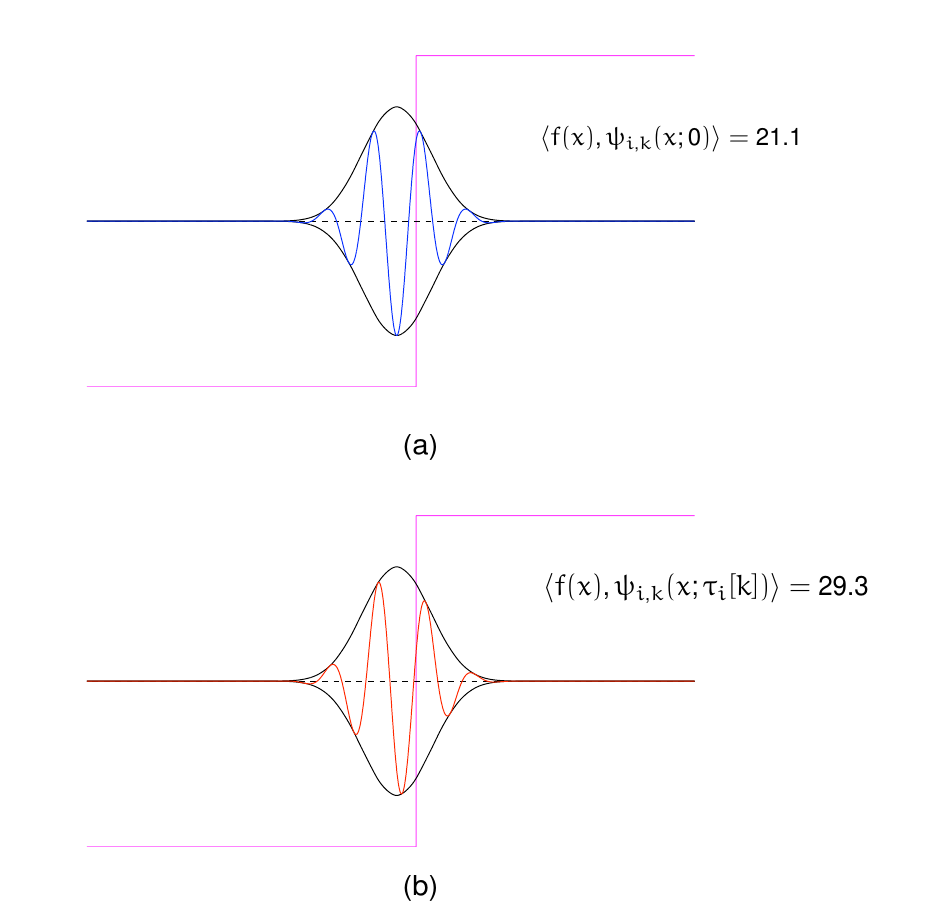} 
\caption{Wavelets corresponding to the step unit $f(x)$: (a) reference wavelet $\psi_{i,k}(x)=\psi_{i,k}(x;0)$ corresponding to the conventional DWT;  (b) shifted wavelet dual-tree $\psi_{i,k}(x;\tau_i[k])$. The magnitudes of the signal correlation in either case clearly shows that the shifted wavelet has a better lock on the singularity.}
\label{fig4}
\end{figure*}

	In figure \ref{fig4}, we demonstrate the shiftability of the dual-tree transform using a Gabor-like B-spline wavelet $\psi(x)$ of degree 3 as the reference. To this end, we consider the step input $f(x)=\sign(x-x_0)$ that has a discontinuity at $x_0$. The $M$-level decomposition of this signal (cf. \cite{kunal_journal} for implementation details) in terms of the conventional DWT is given by
\begin{equation}
\label{exp1}
f(x)=\sum_{\ell} p_{\ell} \varphi_{M,\ell}(x)+\sum_{i=1}^M \sum _k a_{i,k} \psi_{i,k}(x),
\end{equation}
with $\varphi_{M,\ell}(x)$ denoting the translates of the coarse representation of the scaling function. As for the DT-$\mbb{C}$WT, we have the representation
\begin{align}
\label{exp2}
f(x)&= \sum_{\ell} p_{\ell} \varphi_{M,\ell}(x)+\sum_n p'_n \varphi'_{M,n}(x)+\sum_{i=1}^M \sum _k   |c_i[k]| \ \psi_{i,k} \big(x;\tau_i[k]\big).
\end{align}
The idea here is to demonstrate that the shifted wavelets in \eqref{exp2} respond better to the signal transition than the wavelets in \eqref{exp1}; that is, the oscillations of the shifted wavelets have a better lock on the singularity at $x_0$. Figure \ref{fig4} shows the reference wavelet $\psi_{i,k}(x)$ and the shifted wavelet $\psi_{i,k} (x;\tau_i[k])$ corresponding to a specific resolution $i=J$ and translation $k =[x_0/2^J]$ (around the position of the singularity at the coarser resolution). Also shown in the figures are the step input and the fixed Gaussian-like localization window of the wavelet. The oscillation of the shifted wavelet is clearly seen to have a better lock on the transition than the reference.  
The magnitude of the signal correlation in either case justifies this observation as well.

\subsection{Quality Metrics for Dual-Tree Wavelets}
\label{metrics}
 
\begin{table}[!t]
\renewcommand{\arraystretch}{1.3}
\caption{Quality metrics for different dual-tree wavelets} 
\label{table_metrics}
\vspace{0.2cm}
\centering
\begin{tabular}{|c|c|c|}
\hline
Type of Dual-Tree Wavelets & $\varrho$  & $\kappa$ \\
\hline 
Shannon wavelets (ideal) & $1$ & $0$ \\
B-spline wavelets, degree=$1$ & $1$ & $0.9245$ \\
B-spline wavelets, degree=$3$ & $1$ & $0.0882$ \\
B-spline wavelets, degree=$6$ & $1$ & $0.0373$ \\
Orthonormal spline wavelets, degree=$1$ & $1$ & $0.9292$ \\
Orthonormal spline wavelets, degree=$3$ & $1$ & $0.1570$ \\
Orthonormal spline wavelets, degree=$6$ & $1$ & $0.0612$ \\
Kingsbury's wavelets (q-shift Le Gall 5/3) \cite{kingsbury2} & $0.9992$ & $0.6586$ \\
\hline
\end{tabular}
\end{table}

	 The shiftability of the DT-$\mbb{C}$WT was established based on two fundamental properties of the wavelets, $\psi_1(x)$ and $\psi_2(x)$, of the two branches:\\ 
(C$1$) HT correspondence:  $\psi_2(x)=\Hil \psi_1(x)$, \\
(C$2$) Their modulated forms: $\psi_1(x)=\varphi(x)\cos(\w_0x+\xi_0)$, and $\psi_2(x)=\varphi(x)\sin(\w_0x+\xi_0)$, with the support of $\hat\varphi(\w)$ restricted to $[-\w_0,\w_0]$.
\vspace{0.05in}

  The practical challenge is the construction of different flavors of wavelets that fulfill or, at least, provide close approximations of these criteria. Indeed, the first criteria has been marked by an extensive research into the problem of designing both approximate and exact HT wavelet-pairs \cite{CTDWT,kunal_journal}. 
  
  We propose new design metrics for assessing the quality of the approximation. A simple measure for criterion (C$1$) is the correlation
\begin{equation}
\label{rho_def}
\varrho=\max \left(\frac{\langle \psi_2, \Hil \psi_1 \rangle}{|| \psi_1 || \cdot  || \psi_2 ||} \ , \ 0\right). 
\end{equation}
The Cauchy-Schwarz inequality asserts that $0 \leqslant \varrho \leqslant  1$, where $\varrho=1$ if and only if $\psi_2(x)=\Hil \psi_1(x)$. Thus, the higher the value of $\rho$, the better would be the approximation.

Next, note that criterion (C$2$) also has a simple Fourier domain characterization:  
\begin{equation*}
S(\w)=\hat\psi_1(\w)+j\hat \psi_2(\w) = \begin{cases}  g(\w),  &   0 < \w < \infty,   \\
0,  &  -\infty < \w \leqslant 0. 
\end{cases}\
\end{equation*}
where $g(\w)$ is complex-valued in general, and has a local axis of symmetry within its support. In particular, if $g(\w)$ is constrained to be real-valued (corresponding to a symmetric  $\varphi(x)$), then $\psi_1(x)+j\psi_2(x)$ has a constant instantaneous frequency (derivative of the phase) over its support. A reasonable quality metric for (C$2$) would then be the variation of the instantaneous frequency. Alternatively, we can also assess  the degree of symmetry of $S(\w)$. In particular, we propose the following measure: 
\begin{equation}
\label{kappa_def}
\kappa=\frac{\int_0^{\infty} |S^{\ast}(\bar{\w}+\w)-S(\bar{\w}-\w)| \ d\w}{\int_{-\infty}^{\infty} |S(\w)| \ d\w},
\end{equation}
where $\bar{\w}$, the centroid of $|S(\w)|$, is specified as
\begin{equation*}
\bar{\w}=\frac{\int_{-\infty}^{\infty} \w \ |S(\w)| d\w}{\int_{-\infty}^{\infty}  |S(\w)| d\w}.
\end{equation*}
That is, $\kappa$, which lies between $0$ and $1$, measures the disparity between $S(\w)$ and its reflection around the centroid. Indeed, $\kappa$ equal zero if and only if $S(\w)$ is symmetric (with $\bar{\w}$ as the centre of symmetry). Conversely, a high value of $\kappa$ signifies greater local asymmetry in $S(\w)$, and hence a poorer approximation of the modulation criterion.

 We computed the metrics $\varrho$ and $\kappa$ for different dual-tree wavelets (cf. Table \ref{table_metrics}). The wavelets $\psi_1(x)$ and $\psi_2(x)$ were synthesized using the iterated filterbank algorithm, and the integrals involved in \eqref{rho_def} and \eqref{kappa_def} were realized using high-precision numerical integration. It can easily be verified that $\varrho$ and $\kappa$ are appropriately normalized in the sense that they are invariant to the scale and translation of the synthesized wavelets. This is a necessary criteria since the wavelets are essentially synthesized by the filterbank algorithm at some arbitrary scale and translation.
 
 The spline wavelets are analytic by construction \cite{kunal_journal}, and hence $\varrho=1$ irrespective of their degree. This is a necessary criteria since the synthesised wavelets are  
essentially at some arbitrary scale and translation. However, as their degree increases, the B-spline (resp. orthonormal) wavelets converge to the real and imaginary components of the complex Gabor (resp. Shannon) wavelet which exhibits a symmetric spectrum (see \S \ref{spline} for details). The rapid decrease in the value of $\kappa$ reflects this improvement in symmetry (and also the rate of convergence).    

\section{MULTI-DIMENSIONAL EXTENSION}
\label{directional_wavelets}

		In this section, we extend the amplitude-phase representation derived in \S \ref{shiftability} to the multi-dimensional setting. The key ideas carry over directly,  and the final expressions (cf. \eqref{2D_amp_phase} and \eqref{WFA}) are as simple as their $1$D counterparts. The attractive feature of the multi-dimensional dual-tree wavelets is that, besides improving on the shift-invariance of the corresponding transform, they exhibit better directional selectivity than the conventional tensor-product (separable) wavelets \cite{CTDWT}. For the sake of simplicity, and without the loss of generality, we derive the amplitude-phase representation for the particular two-dimensional setting.  \\

\textbf{Two-Dimensional Dual-Tree Wavelets}: To set up the wavelet notations, we briefly recall the construction framework proposed in \cite{kunal_journal} involving the tensor-products of one-dimensional analytic wavelets. Specifically, let $\varphi(x)$ and $\varphi'(x)$ denote the scaling functions associated with the analytic wavelet $\psi_a(x)=\psi(x)+j\psi'(x)$, where $\psi'(x)=\Hil \psi(x)$. The dual-tree construction then hinges on the identification of four separable multiresolutions of $\mathrm{L}^2(\mathbf{R}^2)$ that are naturally associated with the two scaling functions: the approximation subspaces $V(\varphi)\otimes V(\varphi), V(\varphi) \otimes V(\varphi'), V(\varphi') \otimes V(\varphi)$ and $V(\varphi') \otimes V(\varphi')$, and their multiscale counterparts\footnote{The tensor-product $V(\varphi) \otimes V(\varphi)$ denotes the subspace spanned by the translated functions $\varphi(\cdot-m)\varphi(\cdot-n), (m,n) \in \mathbf{Z}^2$.}. The corresponding separable wavelets -- the `low-high', `high-low' and `high-high' wavelets -- are specified  by
\begin{alignat}{2}
\label{separable_wavelets}
\bar{\psi}_1(\x)&=\varphi(x)\psi(y), &  \hspace{5mm}  \bar{\psi}_{4}(\x)&=\varphi(x)\psi'(y),  \nonumber \\
\bar{\psi}_{2}(\x)&=\psi(x)\varphi(y),  & \hspace{5mm}  \bar{\psi}_{5}(\x)&=\psi(x)\varphi'(y),  \nonumber \\
\bar{\psi}_{3}(\x)&=\psi(x)\psi(y), &  \hspace{5mm}  \bar{\psi}_{6}(\x)&=\psi(x)\psi'(y),  \nonumber \\  \nonumber \\
\bar{\psi}_{7}(\x)&=\varphi'(x)\psi(y) , & \hspace{8mm}  \bar{\psi}_{10}(\x)&=\varphi'(x)\psi'(y), \nonumber \\ 
\bar{\psi}_{8}(\x)&=\psi'(x)\varphi(y), & \hspace{8mm} \bar{\psi}_{11}(\x)&=\psi'(x)\varphi'(y), \nonumber \\ 
\bar{\psi}_{9}(\x)&=\psi'(x)\psi(y), & \hspace{8mm}  \bar{\psi}_{12}(\x)&=\psi'(x)\psi'(y),
\end{alignat}
whereas the dual wavelets $\tilde{\bar{\psi}}_1,\ldots,\tilde{\bar{\psi}}_{12}$ are specified in terms of $\tilde \psi(x)$ and $\tilde \psi'(x)$ (here $\x=(x,y)$ denotes the planar coordinates). As far as the identification of the complex wavelets is concerned, the main issue is the poor directional selectivity of the `high-high' wavelets along the diagonal directions. This problem can, however, be mitigated by appropriately exploiting the one-sided spectrum of the analytic wavelet $\psi_a(x)$, and, in effect, by appropriately combining the wavelets in \eqref{separable_wavelets}. In particular, the complex wavelets
\begin{align}
\label{def_CW}
\Psi_1(\x) &= \psi_a(x) \varphi(y) \ =\bar{\psi}_2(\x)+j\bar{\psi}_8(\x) ,    \nonumber \\
\Psi_2(\x) &= \psi_a(x)  \varphi'(y)=\bar{\psi}_5(\x)+j \bar{\psi}_{11}(\x),   \nonumber \\
\Psi_3(\x) &= \varphi(x) \psi_a(y) \ =\bar{\psi}_{1}(\x)+j\bar{\psi}_{4}(\x),     \nonumber \\
\Psi_4(\x) &= \varphi'(x) \psi_a(y)=\bar{\psi}_{7}(\x)+j\bar{\psi}_{10}(\x),   \nonumber \\
\Psi_5(\x) &= \frac{1}{\sqrt{2}}\psi_a(x)  \psi_a(y) = \left(\frac{\bar{\psi}_{3}(\x)-\bar{\psi}_{12}(\x)}{\sqrt{2}}\right)+j\left(\frac{\bar{\psi}_{6}(\x)+\bar{\psi}_{9}(\x)}{\sqrt{2}}\right), \nonumber \\
\Psi_6(\x) &= \frac{1}{\sqrt{2}}\psi^{\ast}_a(x)  \psi_a(y) =\left(\frac{\bar{\psi}_{3}(\x)+\bar{\psi}_{12}(\x)}{\sqrt{2}}\right)+j\left(\frac{\bar{\psi}_{6}(\x)-\bar{\psi}_{9}(\x)}{\sqrt{2}}\right)
\end{align}
exhibit the desired directional selectivity along the primal orientations $\theta_1=\theta_2=0$, $\theta_3=\theta_4=\pi/2$, $\theta_5=\pi/4$, and $\theta_6=3\pi/4$ respectively \cite{kunal_journal}. The dual complex wavelets $\tilde \Psi_1,\ldots,\tilde \Psi_6$, specified in an identical fashion using the dual wavelets $\tilde{\bar{\psi}}_{p}(\x)$, are also oriented along the same set of directions. \newline

	  \textbf{Directional Hilbert Transform} (dHT):  Having recalled the complex wavelet definitions, we next recall the ``quadrature'' correspondence between the components of the complex wavelets that provides further insight into their directional selectivity. Akin to the HT correspondence, the components can be related through the directional HT $\Hil_{\theta}, 0 \leqslant \theta < \pi$:
\begin{equation*}
\Hil_{\theta} f(\x) \stackrel{\mathscr{F}}{\longleftrightarrow} -j\sign(\u_{\theta}^T \bw)\hat{f}(\bw)
\end{equation*}
where $\u_{\theta}=(\cos \theta,\sin \theta)$ denotes the unit vector\footnote{Note that the half-spaces $\{\bw: \u_{\theta}^T \bw > 0\}$ and $\{\bw: \u_{\theta}^T \bw < 0\}$ play a role, analogous to that played by the half-lines $0  \leqslant w  \leqslant \infty$ and $-\infty \leqslant w \leqslant 0$ in case of the HT, in specifying the action of dHT.} along the direction $\theta$. In particular, one has the correspondences
\begin{equation*}
\mathfrak{Im} (\Psi_{\ell})= \mathcal{H}_{\theta_{\ell}} \mathfrak{Re}(\Psi_{\ell}) \quad (\ell=1,\ldots,6),
\end{equation*}
so that, by denoting the real component of the complex wavelet $\Psi_{\ell}(\x)$ by $\psi_{\ell}(\x)$, we have the following convenient representations
\begin{equation}
\label{analytic_form}
\Psi_{\ell}(\x)=\psi_{\ell}(\x)+j \mathcal{H}_{\theta_{\ell}} \psi_{\ell}(\x)  \quad (\ell=1,\ldots,6)
\end{equation}
which are reminiscent of the $1$D analytic representation.

\subsection{Amplitude-Phase Representation}

	Let us denote the dilated-translated copies of the each of the six analysis wavelets $\tilde\Psi_{\ell}(\x)$ by $\tilde \Psi_{\ell,i,\k}(\x)$, so that
\begin{equation*}
\tilde \Psi_{\ell,i,\k}(\x)=\Xi_{i,\k} \tilde\Psi_{\ell}(\x) \quad  (i \in \mathbf{Z}, \k \in \mathbf{Z}^2),
\end{equation*}
where $\Xi_{i,\k}$ is specified by $\Xi_{i,\k}f(\x)=2^i f(2^i \x-\k)$. The corresponding dual-tree transform involves the analysis of a finite-energy signal $f(\x)$ in terms of the sequence of projections
\begin{equation}
\label{coeff_complex}
 c^{\ell}_i[\k]= \frac{1}{4} \big \langle f, \tilde \Psi_{\ell, i, \k} \big \rangle,
\end{equation}
where the use of the normalization factor $1/4$ will be justified shortly. Before deriving the representation of $f(\x)$ in terms of the analysis coefficients $c^{\ell}_i[\k]$, we introduce the following fractional extensions of the dHT:
\begin{equation}
\label{def_fdHT}
\H_{\theta, \tau}=\cos(\pi\tau)\ \I - \sin(\pi\tau) \ \Hil_{\theta} \quad (\tau \in \mathbf{R})
\end{equation}
that formally allow us to capture the notion of direction-selective phase-shifts. Certain key properties of the fHT carry over to the fdHT. In particular, the family of fdHT operators $\{\H_{\theta, \tau}\}_{\tau \in \mathbf{R}},\ 0 \leqslant \theta <\pi$, exhibit the  fundamental invariances of     
\begin{itemize}
\item Translation: $\H_{\theta,\tau}\{f( \cdot- \y)\}(\x)=(\H_{\theta,\tau} f)(\x - \y)$;
\item Dilation: $\H_{\theta,\tau}\{f(\lambda \cdot)\}(\x)=(\H_{\theta,\tau} f)(\lambda \x)$; and
\item Norm:  $||\H_{\theta,\tau} f ||=||f||,$ for all $f \in \mathrm{L}^2(\mathbf{R}^2)$,
\end{itemize}
which played a decisive role in establishing \eqref{amp-phase}. Based on the above invariances, we can derive the following representation
\begin{equation}
\label{2D_amp_phase}
f(\x)= \sum_{(\ell,i,\k)} \big|c^{\ell}_i[\k] \big|  \ \Xi_{i,\k} \big\{\psi_{\ell} \big(\x ; \tau^{\ell}_i[\k]\big)\big\}
\end{equation}
involving the superposition of direction-selective synthesis wavelets affected with appropriate phase-shifts (cf. Appendix \S\ref{A2} for derivation details). In particular, the wavelets $\psi_{\ell}\big(\x ; \tau^{\ell}_i[\k]\big)$ are derived from the reference wavelet $\psi_{\ell}(\x)$ through the action of fdHT, corresponding to the direction $\theta_{\ell}$ and shift $\tau^{\ell}_i[\k]=\arg(c^{\ell}_i[\k])/\pi$ As in the $1$D setting, further insight into the above representation is  obtained by considering wavelets resembling windowed plane waves.

\subsection{Directional Modulated Wavelets}

A distinctive feature of the dHT is its phase-shift action in relation to plane-waves: it transforms the directional cosine $\cos(\u_{\theta}^T\x)$ into the directional sine $\sin(\u_{\theta}^T\x)$. Moreover, what turns out to be even more relevant in the current context is that the above action is preserved for certain classes of windowed plane waves; in particular, we have 
\begin{equation*}
\Hil_{\theta} \left\{ \varphi(\x) \cos(\Omega \u_{\theta}^T\x) \right\}=\varphi(\x) \sin(\Omega \u_{\theta}^T\x)
\end{equation*}
provided that $\varphi(\x)$ bandlimited to the disk $D_{\Omega}=\{ \bw: ||\bw|| < \Omega\}$. The following result -- a specific multi-dimensional extension of Theorem \ref{Bedrosian} -- then follows naturally for the fractional extensions:
\vspace{0.05in}	
\begin{proposition} Let the window function $\varphi(\x)$ be bandlimited to the disk $D_{\Omega}$. Then we have that
\begin{equation}
\label{modulation_fdHT}
\H_{\theta,\tau} \left\{ \varphi(\x) \cos(\Omega \u_{\theta}^T\x) \right\}=\varphi(\x) \sin(\Omega \u_{\theta}^T\x+\pi \tau).
\end{equation}
\end{proposition}
\vspace{0.05in}	
That is, the fdHT acts only on the phase of the oscillation while the window remains fixed. In particular, if the dual-tree wavelets are of the form       
\begin{equation}
\label{law1}
\psi_{\ell}(\x) = \upvarphi_{\ell}(\x) \cos\left(\Omega_{\ell} \u_{\theta_{\ell}}^T\x\right), \quad \ell=1,\ldots, 6,
\end{equation}
then the right-hand side of \eqref{2D_amp_phase} assumes the form 
\begin{equation*}
\sum_{(\ell,i,\k)} \stackrel{\mathrm{fixed \ window}}{\overbrace{\upvarphi_{\ell, i,\k}(\x)}}  \Xi_{i,\k}  \Big\{\stackrel{\mathrm{variable \ amp-phase \ directional \ wave}}{ \overbrace{|c^{\ell}_i[\k]| \cos\left(\Omega_{\ell} \u_{\theta_{\ell}}^T\x+\pi \tau^{\ell}_i[\k]\right)}} \Big\},
\end{equation*}
where $\upvarphi_{\ell, i,\k}(\x)$ are the dilated-translated copies of the window $\varphi_{\ell}(\x)$. The above expression explicitly highlights the role of $\tau^{\ell}_i[\k]$ as a scale-dependent measure of the local signal displacements along certain preferential directions.

	Indeed, this is the scenario for the spline-based transform proposed in \cite{kunal_journal} where the dual-tree wavelets asymptotically converge to the two-dimensional Gabor functions proposed by Daugman \cite{discrete_Gabor}. Moreover, these Gabor-like dual-tree wavelets were constructed using the B-spline scaling function and the semi-orthogonal B-spline wavelet. Replacing these with the orthonormal B-spline and the orthonormal wavelet, respectively, would then result in Shannon-like dual-tree wavelets -- $\mathrm{sinc}$-windowed directional plane waves -- following the fact that the orthonormal spline multiresolution asymptotically converges to the Shannon multiresolution (cf. \S\ref{mod_wavelets}).   flip the roles of the analysis and synthesis wavelets: we analyze the signal using the dual complex wavelet $\tilde \Psi(x;\alpha)=\tilde \psi(x;\alpha)+j \tilde \psi'(x;\alpha)$, and the Gabor-like wavelet $\Psi(x;\alpha)$ is used for reconstruction.

 \section{SHIFTABLE SPLINE WAVELETS}
\label{spline}

	If the wavelet is not modulated, we can characterize the action of the fHT and the fdHT by studying the family of wavelets $\{\psi(x; \tau)\}_{\tau \in \mathbf{R}}$ and $\{\psi_{\ell}(\x; \tau)\}_{\tau \in \mathbf{R}}$, respectively. The remarkable fact is that it can be done explicitly for all spline wavelets derived from the fractional B-splines \cite{alpha-tau}, which are the fractional extensions of the polynomial B-splines. \\
	
	\textbf{Spline Multiresolution}:  We recall that the B-spline $\beta_{\tau}^{\alpha}(x)$, of degree $\alpha \in \mathbb{R}^{+}_{0}$ and shift $\tau \in \mathbb{R}$, is specified via the Fourier transform
\begin{equation}
\label{def_Bspline}
\beta_{\tau}^{\alpha}(x) \stackrel{\mathscr{F}}{\longleftrightarrow}  \left( \frac{1-e^{-j\omega}}{j\omega}\right)^{\frac{\alpha+1}{2}+\tau} \left( \frac{1-\mathrm{e}^{j\omega}}{-j\omega}\right)^{\frac{\alpha+1}{2}-\tau}.
\end{equation} 
The degree primarily controls the width (and norm) of the function, whereas the shift influences the phase of the Fourier transform. As will be seen shortly, the latter property plays a key role in conjunction with the fHT. Though these functions are not compactly supported in general, their $O(1/|x|^{\alpha+2})$ decay ensures their inclusion in $\mathrm{L}^1(\mathbb{R}) \cap \mathrm{L}^2(\mathbf{R})$. More crucially, they satisfy certain technical criteria \cite{alpha-tau,FS}  needed to generate a valid multiresolution of $\mathrm{L}^2(\mathbf{R})$: \newline
\vspace{0.05in}	
(MRA$1$) Riesz-basis property: the subspace 
\begin{equation*}
\mathrm{span}_{\ell^2} \{\beta_{\tau}^{\alpha}(\cdot-k)\}_{k \in \mathbb{Z}}
\end{equation*}
admits a stable Riesz basis. \newline
(MRA$2$) Two-scale relation; the refinement filter is specified by the transfer function
\begin{equation*}
H^{\alpha}_{\tau}(\mathrm{e}^{j\omega})= \frac{1}{2^{(\alpha+1)}} \left(1+e^{-j\omega}\right)^{\frac{\alpha+1}{2}+\tau}  \left(1+\mathrm{e}^{j\omega}\right)^{\frac{\alpha+1}{2}-\tau}.
\end{equation*} 
(MRA$3$) Partition-of-unity property. 
\vspace{0.05in}

	 \textbf{Spline Wavelets}: As far as the wavelet specification is concerned, the transfer function of the wavelet filter that generates the generic spline wavelet $\psi_{\tau}^{\alpha}(x)$, of degree $\alpha$ and shift $\tau$, is given by
\begin{equation}
\label{wavelet_filter}
G^{\alpha}_{\tau}(\mathrm{e}^{j\omega})=\mathrm{e}^{j\omega} Q^{\alpha}(-e^{-j\omega})H^{\alpha}_{\tau}(-e^{-j\omega},
\end{equation}
where the filter $Q^{\alpha}(\mathrm{e}^{j\omega})$ satisfies the lowpass constraint $Q^{\alpha}(1)=1$, and is independent of $\tau$. The filter plays a crucial role in differentiating between various orthogonal (e.g., Battle-Lemarié wavelet) and biorthogonal (e.g., semi-orthogonal B-spline wavelet) flavors of spline wavelets of the same order \cite{convergence,semi-ortho}. The associated dual multiresolution is specified by a dual spline function and a dual spline wavelet $\tilde{\psi}_{\tau}^{\alpha}(x)$; the fundamental requirement for the dual system is that the dual wavelet -- generated using a transfer function similar to that in \eqref{wavelet_filter} -- satisfies the biorthogonality criterion $\langle {\psi}_{\tau}^{\alpha}(\cdot -m),  \tilde {\psi}_{\tau}^{\alpha}(\cdot-n)\rangle=\delta[m-n]$. We shall henceforth use the notation $\psi_{\tau}^{\alpha}(x)$ to denote a spline wavelet of order $\alpha$ and shift $\tau$, irrespective of its genus (orthonormal, B-spline, dual spline, etc.).

 It turns out that the family of spline wavelets $\{\psi_{\tau}^{\alpha}\}_{\tau \in \mathbf{R}}$ (of a specified genus and order) is closed with respect to the fHT operation.
	
\vspace{0.05in}	
\begin{proposition} 
\label{fHT_wavelets} 
The fHT of a spline wavelet is a spline wavelet of same genus and order, but with a different shift. In particular, 
 \begin{equation*}
\H_{\bar \tau} \psi_{\tau}^{\alpha}(x)=\psi_{\tau-\bar \tau}^{\alpha}(x).
\end{equation*}
\end{proposition}
\vspace{0.05in}	

 	The above result (cf. part I of Appendix \S\ref{A3}) signifies that the fHT acts only on the shift parameter of the spline wavelet while preserving its genus and order. Thus, for the dual-tree transform involving the corresponding wavelet basis $\{\psi_{i,k}^{\alpha}(x)\}_{(i,k) \in \mathbf{Z}^2}$ and its HT pair, we have the following signal representation   
\begin{align*}
f(x)&= \sum_{(i,k)}  |c_i[k]| \ \Xi_{i,k} \big \{\H_{\tau_i[k]} \psi^{\alpha}_{\tau}(x) \big \}  \nonumber \\
&= \sum_{(i,k)}   \ \Xi_{i,k} \big \{ |c_i[k]|\  \psi^{\alpha}_{\tau-\tau_i[k]}(x) \big \}
\end{align*}
involving the weighted sum of the appropriately ``shifted'' spline wavelets.

	Finally, we investigate the action of the fdHT on the $2$D dual-tree wavelets $\psi_{\ell}(\x)$ constructed using a spline wavelet $\psi^{\alpha}_{\tau}(x)$ of a specific genus, and its HT pair $\psi^{\alpha}_{\tau+1/2}(x)$ \cite{kunal_journal}. It turns out that, as in the $1$D case, the action is purely determined by the perturbation of the shift parameter of the constituent spline functions. However, the key difference is that the fdHT operators act ``differentially'' on the shifts of the spline functions along each dimension. Before stating the result we briefly digress to introduce a convenient notation. Observe that the six dual-tree wavelets are of the general form
 \begin{equation*}
 \sum_{\ell} g_{\ell}(x; \alpha, \tau_x) h_{\ell}(y; \alpha, \tau_y)
\end{equation*}
where $g_{\ell}(x; \alpha, \tau_x)$ and $h_{\ell}(y; \alpha, \tau_y)$ are spline scaling functions/wavelets that have a common degree $\alpha$ but whose shifts depend on $\tau_x$ and $\tau_y$respectively. To explicitly emphasize the dependence on the parameters $\tau_x$ and $\tau_y$, we denote  the dual-tree wavelets by $\psi_{\ell}(\x;\alpha,\boldsymbol \tau)$ with the shift-vector $\btau=(\tau_x,\tau_y)$ specifying the shift parameters of the spline functions involved along each dimension. For instance, the wavelets  $\psi_1(\x;\alpha,\boldsymbol \tau)$ and $\psi_5(\x;\alpha,\boldsymbol \tau)$ are specified (see \eqref{def_CW}) by	 
\begin{align*}
\psi_1(\x;\alpha,\boldsymbol \tau)&= \psi^{\alpha}_{\tau_x}(x)\beta^{\alpha}_{\tau_y}(y),  \nonumber \\
\psi_5(\x;\alpha,\boldsymbol \tau)&=\frac{1}{\sqrt 2}\big(\psi^{\alpha}_{\tau_x}(x)\psi^{\alpha}_{\tau_y}(y)-\psi^{\alpha}_{\tau_x+1/2}(x)\psi^{\alpha}_{\tau_y+1/2}(y)\big),
\end{align*}
where $\btau=(\tau,\tau)$ by construction. In general, setting $\btau=(\tau,\tau)$ for all the six wavelets we deduce (see part II of Appendix \S\ref{A3} for a proof) the following:

\vspace{0.05in}	
\begin{proposition}
\label{fdHT_splinewavelets} The fdHT of a $2$D dual-tree spline wavelet is a dual-tree spline wavelet of the same order and direction, but with a different shift:  
\begin{equation*}
\H_{\theta_{\ell},\bar \tau} \psi_{\ell}(\x;\alpha,\btau) = \psi_{\ell}\big(\x; \alpha, \btau -  \bar\tau \mu_{\ell}  \boldsymbol u_{\theta_{\ell}}\big)
\end{equation*}
where $\mu_{\ell}=1$ for $\ell=1,\ldots,4$, and equals $1/\sqrt 2$ for $\ell=5$ and $6$.
\end{proposition}
\vspace{0.05in}

	The result is quite intuitive. The horizontal and vertical wavelets can be `shifted' along the direction of the corresponding fdHT by perturbing the shift of the spline functions running along the same direction; the shift of the spline functions along the orthogonal direction remains unaffected. However, the diagonal wavelets can be `shifted' only by simultaneously by perturbing the shift of the splines along both dimensions. 

	Thus, as a direct consequence of  \eqref{2D_amp_phase} and proposition \ref{fdHT_splinewavelets}, we have the following signal representation: 
\begin{align*}
f(\x)=\sum_{\ell,i,\k}  \Xi_{i,\k}   \Big \{ \big |c^{\ell}_i[\k] \big| \ \psi_{\ell} \big(\x; \alpha, \btau - \mu_k  \tau^{\ell}_i[\k]   \boldsymbol u_{\theta_{\ell}}\big) \Big \}
\end{align*}
where the shift information $\tau^{\ell}_i[\k]=\arg(c^{\ell}_i[\k])$, at different scales along each of the six directions, is directly encoded into the shift-parameter of the spline wavelets. As discussed in \S\ref{directional_wavelets}, for sufficiently large $\alpha$, $\psi_{\ell}(\x;\alpha,\boldsymbol \tau)$ constructed using the B-spline (orthonormal spline) wavelets resemble the Gabor (resp. Shannon) wavelet where the shift $\tau^{\ell}_i[\k]$ gets directly incorporated into the phase of the modulating plane wave.

\section{CONCLUDING REMARKS}
\label{conclusion}

	We derived an insight into the improved shift-invariance of the dual-tree complex wavelet transform based on single fundamental attribute of the same: the HT correspondence between the wavelet bases. Indeed, the identification of the fHT-transformed wavelets involved in representation \eqref{WFA} followed as a direct consequence of this correspondence. The shiftability of the transform was then established based on two key results:
\begin{itemize}
\item the intrinsic invariances of the fHT group with respect to translations, dilations and norm-evaluations; and
\item theorem \eqref{Bedrosian} describing the phase-shift action of the fHT on modulated wavelets.
\end{itemize}
In particular, a multiscale amplitude-phase signal representation was derived for the class of the modulated wavelets which highlighted the additional freedom of the wavelets to lock on to singularities of the signal. We also proposed certain metrics for accessing the modulation criteria and the quality of the HT correspondence between the dual-tree wavelets. These could prove useful in the design of new dual-tree wavelets with better shiftability.
		
	Before concluding, we would like to remark that the (direction-selective) shiftability of the dual-tree transform can also be extended to higher dimensions. In particular, the wavelet construction \eqref{def_CW}, the fHT correspondences \eqref{analytic_form}, the modulation law \eqref{modulation_fdHT}, and, crucially, the amplitude-phase representation \eqref{2D_amp_phase} carry over directly to the multi-dimensional setting.

\section{Appendix}

 \subsection{Proof of Theorem \ref{representation}}
 \label{A1} 
 
The sufficiency part of the theorem follows from the properties of the fHT operator; we only need to prove the converse. It is well-known that a unitary linear operator $T$ on $\mathrm{L}^2(\mathbf{R})$ is translation invariant if and only if there exists a bounded (complex-valued) function $m(\omega)$ such that 
 \begin{equation*}
 \widehat{Tf}(\w)=m(\w)\hat f(\w), 
\end{equation*}
for all $f \in \mathrm{L}^2(\mathbf{R})$ \cite[Chapter 1]{Stein_Weiss}. This Fourier domain characterization reduces the problem to one of specifying a bounded function $m(\w)$ such that $T$ has the desired invariances. It can be readily demonstrated that the dilation-invariance criterion translates into the constraint 
\begin{equation}
\label{scaling}
m(a\w)=m(\w), \  a>0.
\end{equation}
Moreover, since the real and imaginary components of $m(\w)$ must independently satisfy \eqref{scaling}, one even has $m_1(a\w)=m_1(\w)$ and $m_2(a\w)=m_2(\w)$,  for all $a>0$, where $m_1(\w)$ and $m_2(\w)$ are the real and imaginary components of $m(\w)$.

	Next, observe that the Hermitian symmetry requirement 
\begin{equation}
\label{Hermititan}
m^{\ast}(\w)=m(-\w) 
\end{equation}
on the multiplier require $m_1(\w)$ and $m_2(\w)$ to be even and odd symmetricrespectively; that is, $m_1(-\w)=m_1(\w)$ and $m_2(-\w)=-m_2(\w)$. These constitute the crucial relations, since one can easily verify that the only bounded functions (up to a scalar multiple) that satisfy \eqref{scaling} and \eqref{Hermititan} simultaneously are the constant function $m_1(\w) \equiv 1$, and its ``skewed'' counterpart $m_2(\w)=\sign(\w)$; that is, it is both necessary and sufficient that 
\begin{equation*}
m(\w)=\gamma_1+j \gamma_2 \sign(\w),
\end{equation*}
for some real $\gamma_1$ and $\gamma_2$. Finally, combining the equivalence 
\begin{align*}
||Tf||^2= \frac{1}{2\pi} \int |m(\w)|^2 |\hat f(\w)|^2 \mathrm{d}\w = (\gamma_1^2+\gamma_2^2)^2 ||f||^2
\end{align*}
obtained through Parseval's identity, with the norm invariance requirement we arrive at the criterion $\gamma_1^2+\gamma_2^2=1$. Therefore, it is both necessary and sufficient that $m(\w)=\cos\theta+j \sin \theta \ \sign(\w)$ for some $\theta \in (-\pi,\pi]$, and this establishes the representation
\begin{equation*}
T=\cos \theta \ \I-\sin \theta \ \Hil.
\end{equation*}
The uniqueness of $\theta \in (-\pi,\pi]$ necessarily follows from the quadrature correspondence in \eqref{quadrature_corresp}. $\quad \Box$

 \subsection{Proof of Theorem \ref{Bedrosian}}
 \label{AA}

We note that the Fourier transform of $f(x)g(x)$ is given by the convolution $(2\pi)^{-1}(\hat f \ast \hat g)(\w)$. Thus, if we denote the Fourier transform of $\H_{\tau} \big \{f(x)g(x) \big \}$ by $F(\w)$, then following definition \eqref{freq_response}, we have that 
\begin{align*}
F(\w)&=\mathrm{e}^{j\pi \tau \sign(\w)} \widehat {(f  g)}(\w) = \mathrm{e}^{j\pi \tau \sign(\w)}\frac{1}{2\pi}\int_{\mathbf{R}} \hat f(\xi) \hat g(\w-\xi) \mathrm{d}\xi.
\end{align*}
In particular, this gives us the Fourier representation
\begin{align}
\label{expansion_product}
\H_{\tau} \big \{f(x)g(x) \big \}&=  \frac{1}{2\pi} \int_{\mathbf{R}} F(\w)  \mathrm{e}^{j\w x}\mathrm{d}\w  \nonumber \\
&= \frac{1}{2\pi} \int_{\mathbf{R}}  \mathrm{e}^{j\w x}  \mathrm{e}^{j\pi \tau \sign(\w)} \left( \frac{1}{2\pi}\int_{\mathbf{R}} \hat f(\xi) \hat g(\w-\xi)  \mathrm{d}\xi \right)  \mathrm{d}\w.
\end{align}
Now, by commuting the order of the integrals and by applying the frequency translation $\zeta=\w-\xi$, we can rewrite \eqref{expansion_product} as the double integral
\begin{align}
\label{reduction}
&\frac{1}{4\pi^2}  \int_{\mathbf{R}^2}  \mathrm{e}^{j\w x}  \hat f(\xi)    \mathrm{e}^{j\pi \tau \sign(\w)} \ \hat g(\w-\xi) \mathrm{d}\w \mathrm{d}\xi  \nonumber \\
&= \frac{1}{4\pi^2}  \int_{Q}  \mathrm{e}^{j(\xi+\zeta)x}   \hat f(\xi) \  \mathrm{e}^{j\pi \tau \sign(\xi+\zeta)} \  \hat g(\zeta) \mathrm{d}\zeta  \mathrm{d}\xi.
\end{align}
The effective domain of integration in \eqref{reduction} gets restricted to the region $Q=\{(\zeta,\xi):  \ |\zeta| \geqslant \Omega, \  |\xi| < \Omega\}$ as a consequence of the assumptions on the supports of $ \hat f(\xi)$ and $ \hat g(\zeta)$. In particular, it can easily be verified that $ \sign(\xi+\zeta)= \sign(\zeta)$ on $Q$; this allows us to factor \eqref{reduction} into two separate integrals giving the desired result:
\begin{align*}
\H_{\tau} \big \{f(x)g(x) \big \} &= \frac{1}{4\pi^2}  \int_{Q} \mathrm{e}^{j \xi x} \mathrm{e}^{j \zeta x} \hat f(\xi) \ \mathrm{e}^{j\pi \tau \sign(\zeta)} \  \hat g(\zeta) \mathrm{d}\zeta  \nonumber \\
&= \left(\frac{1}{2\pi}  \int_{ |\xi| < \Omega} \mathrm{e}^{j \xi x} \hat f(\xi)\mathrm{d}\xi  \right) \left( \frac{1}{2\pi}  \int_{ |\zeta| \geqslant \Omega}  \mathrm{e}^{j \zeta x} \widehat{(\H_{\tau} g)}(\zeta) \mathrm{d}\zeta \right) \nonumber \\
&=f(x) \H_{\tau}g(x). \quad \Box
\end{align*}

\subsection{Derivation of Equation \eqref{2D_amp_phase}}
\label{A2} 	
 
 	As a first step, we consider the equivalent representations of $f(x)$ in terms of the four distinct bases generated through the dilations-translations of the separable wavelets in \eqref{separable_wavelets}:
\begin{align}
\label{quad_tree}
f(\x)= \sum_{(i,\k)}& \big(a_{1+p,i}[\k] \bar{\psi}_{1+p,i,\k}(\x) + a_{2+p,i}[\k] \bar{\psi}_{2+p,i,\k}(\x) + a_{3+p,i}[\k] \bar{\psi}_{3+p,i,\k}(\x) \big) 
\end{align}
for $p=0,3,6$ and $9$, where the expansion coefficients are specified by 
\begin{equation}
\label{separable_coeff}
a_{q,i}[\k]=\langle f, \widetilde{\bar{\psi}}_{q,i,\k} \rangle \qquad (q=1,\ldots,12).
\end{equation}

Next, we combine and regroup these expansions as follows
\begin{align*}
&f(\x)=  \sum_{(i,\k)}\Bigg\{ \frac{1}{4}\ \Big( a_{2,i}[\k] \bar{\psi}_{2,i,\k}(\x)+ a_{8,i}[\k] \bar{\psi}_{8,i,\k}(\x)\Big)+  \frac{1}{4}\Big( a_{5,i}[\k] \bar{\psi}_{5,i,\k}(\x)+ a_{11,i}[\k] \bar{\psi}_{11,i,\k}(\x)\Big) \nonumber \\
&+  \frac{1}{4} \Big( a_{1,i}[\k] \bar{\psi}_{1,i,[\k]}(\x)+ a_{4,i}[\k] \bar{\psi}_{4,i,\k}(\x)\Big) + \frac{1}{4} \Big( a_{7,i}[\k] \bar{\psi}_{7,i,\k}(x) + a_{10,i}[\k] \bar{\psi}_{10,i,\k}(\x)\Big)  \nonumber \\
&+ \frac{1}{4\sqrt 2}(a_{3,i}[\k]- a_{12,i}[\k]) \left( \frac{\bar{\psi}_{3,i,\k}(\x) - \bar{\psi}_{12,i,\k}(\x)}{\sqrt 2}\right)  + \frac{1}{4\sqrt 2}(a_{6,i}[\k] + a_{9,i}[\k]) \left( \frac{\bar{\psi}_{6,i,\k}(\x)+ \bar{\psi}_{9,i,\k}(\x) }{\sqrt 2}\right) \nonumber \\
&+ \frac{1}{4\sqrt 2}(a_{3,i}[\k] + a_{12,i}[\k]) \left( \frac{\bar{\psi}_{3,i,\k}(\x) + \bar{\psi}_{12,i,\k}(\x) }{\sqrt 2}\right)  + \frac{1}{4\sqrt 2}(a_{6,i}[\k] - a_{9,i}[\k]) \left( \frac{\bar{\psi}_{6,i,\k}(\x) - \bar{\psi}_{9,i,\k}(\x) }{\sqrt 2}\right) \Bigg\}.
\end{align*}
The terms on the right-hand side can now be conveniently expressed in terms of the coefficients $c^{\ell}_i[\k]=\big |c^{\ell}_i[\k]\big| \mathrm{e}^{j\phi^{\ell}_i[\k]}$ and the dual-tree wavelets $\psi_{\ell,i,\k}(\x)$. For instance, consider the terms in the third line. The wavelet pair $\psi_{5,i,\k}(\x)$ and $\H_{\theta_5}\psi_{5,i,\k}(\x)$ are readily identified; moreover, the correspondences 
\begin{align*}
&\frac{(a_{3,i}[\k]- a_{12,i}[\k])}{4 \sqrt 2}=\frac{1}{4}\mathfrak{Re} \langle f, \widetilde \Psi_{5,i,\k} \rangle=|c^5_i[\k]| \cos \phi^5_i[\k];   \nonumber \\
&\frac{(a_{6,i}[\k]+ a_{9,i}[\k])}{4 \sqrt 2} =\frac{1}{4}\mathfrak{Im} \langle f,  \widetilde \Psi_{5,i,\k} \rangle=-|c^5_i[\k]| \sin \phi^5_i[\k],
\end{align*}
that follow from \eqref{coeff_complex} and \eqref{separable_coeff}, allow us to rewrite it as
\begin{align*}
&|c^5_i[\k]| \Big(\cos \phi^5_i[\k] \psi_{5,i,\k}(\x)  - \sin \phi^5_i[\k] \H_{\theta_5}\psi_{5,i,\k}(\x)\Big) =|c^5_i[\k]| \ \psi_{5,i,\k}(\x;\tau^5_i[\k])
\end{align*}
with the shift specified by $\tau^5_i[\k]=\phi^5_i[\k]/\pi$. Subjecting the rest of the terms to a similar 
treatment we arrive at the desired representation.

 \subsection{Proof of Propositions \ref{fHT_wavelets} and \ref{fdHT_splinewavelets}}
 \label{A3} 	

\textbf{Part I}:  This result is easily established using an auxiliary operator. Specifically, we consider the fractional finite-difference (FD) operator
 \begin{equation}
 \label{def_FD}
\Delta_{\tau}f(x)  \stackrel{\mathscr{F}}{\longleftrightarrow} \D_{\tau}(\mathrm{e}^{j\omega}) \hat{f}(\omega),
 \end{equation}
corresponding to the frequency response $\D^{\alpha}_{\tau}(\mathrm{e}^{j\omega})=\left(1-e^{-j\omega}\right)^{\tau} \left(1-\mathrm{e}^{j\omega}\right)^{-\tau}$, which allows us to relate fractional B-splines (and the corresponding filters) of the same order but with different shifts. In particular, it allows us to express the action of the fHT on a B-spline as a linear (digital) filtering:
\begin{align}
\label{fHT_spline}
\H_{\bar \tau} \beta_{\tau}^{\alpha}(x)&=\Delta_{\bar \tau} \beta_{\tau-\bar \tau}^{\alpha}(x)  \nonumber \\
&=\sum_{k \in \mathbf{Z}} \d_{\tau}[k]  \beta_{\tau-\bar \tau}^{\alpha}(x-k).
\end{align}

	Indeed, based on definitions \eqref{def_Bspline} and \eqref{def_FD}, and the identity\footnote{We specify the fractional power of a complex number $z$ by $z^{\gamma}=|z|^{\gamma}\mathrm{e}^{j\gamma \arg(z)}$ corresponding to the principal argument $|\arg(z)| <\pi$. On this branch, the identity $(z_1 z_2)^{\gamma}=z_1^{\gamma}z_2^{\gamma}$ holds only if $\arg(z_1)+ \arg(z_1) \in (-\pi,\pi)$ \cite[Chapter 3]{Stein_ComplexAnalysis}.} $(j\w)^{-\bar \tau} (-j\w)^{\bar \tau}=\exp\big(-j\pi \tau \sign(\w)\big)$, we have the following factorization 
\begin{align*}
\hat \beta_{\tau}^{\alpha}(\w) &= (j\w)^{-\bar \tau} (-j\w)^{\bar \tau} \D_{\bar \tau}(\mathrm{e}^{j\omega}) \hat \beta_{\tau-\bar \tau}^{\alpha}(\w) \nonumber \\
&=\exp\big(-j\pi \bar \tau \sign(\w)\big)  \D_{\bar \tau}(\mathrm{e}^{j\omega}) \hat \beta_{\tau-\bar \tau}^{\alpha}(\w)
\end{align*}
which results in the equivalence
\begin{align*}               
\H_{\bar \tau}  \beta_{\tau}^{\alpha}(x)& \stackrel{\mathscr{F}}{\longleftrightarrow}   \exp\big(j\pi \tau \sign(\w)\big) \hat \beta_{\tau}^{\alpha}(\w) \nonumber \\
& = \D_{\bar \tau}(\mathrm{e}^{j\omega}) \hat \beta_{\tau-\bar \tau}^{\alpha}(\w) \\
& \stackrel{\mathscr{F}}{\longleftrightarrow}  \Delta_{\bar \tau} \beta_{\tau-\bar \tau}^{\alpha}(x). 
\end{align*}
 		
	Next, we observe that the \textit{conjugate-mirrrored} version of the FD filter can also be used to relate the spline refinement (scaling) filters of the same order but with different shifts: $H_{\tau-\bar \tau}^{\alpha}(\mathrm{e}^{j\omega})= \D_{\bar \tau}(-e^{-j\omega})$ $ H_{\tau}^{\alpha}(\mathrm{e}^{j\omega})$. In particular, we have the relation $g_{\tau-\bar \tau}^{\alpha}[k]=(\d_{\bar \tau} \ast g_{\tau}^{\alpha})[k]$ between the corresponding wavelet filters:
\begin{align}
\label{wav_filters}
G_{\tau-\bar \tau}^{\alpha}(\mathrm{e}^{j\omega})&=\mathrm{e}^{j\omega} Q^{\alpha}(-e^{-j\omega})H^{\alpha}_{\tau- \bar \tau}(-e^{-j\omega}) \nonumber   \\
&=  \D_{\bar \tau}(\mathrm{e}^{j\omega}) \mathrm{e}^{j\omega} Q^{\alpha}(-e^{-j\omega}) H_{\tau}^{\alpha}(-e^{-j\omega}) \nonumber \\
&=\D_{\bar \tau}(\mathrm{e}^{j\omega})  G_{\tau-\bar \tau}^{\alpha}(\mathrm{e}^{j\omega}).
\end{align}
As a result of \eqref{fHT_spline} and \eqref{wav_filters}, we have the desired equivalence:
\begin{align*}
\H_{\bar \tau} \psi_{\tau}^{\alpha}(x/2) &=\sum_{k \in \mathbb{Z}} g_{\tau}^{\alpha}[k] \H_{\bar \tau} \beta_{\tau}^{\alpha}(x-k)\nonumber   \\
&=\sum_{k \in \mathbb{Z}} g_{\tau}^{\alpha}[k] \Big\{ \sum_{n \in \mathbb{Z}} d_{\bar \tau}[n] \beta_{\tau-\bar \tau}^{\alpha}(\cdot- n)\Big\}(x-k) \nonumber  \\
&=\sum_{m \in \mathbb{Z}} (g_{\tau}^{\alpha} \ast d_{\bar \tau})[m] \beta_{\tau-\bar \tau}(x-m) \nonumber  \\
&= \psi_{\tau- \bar \tau}^{\alpha}(x/2).  \quad  
\end{align*}

\textbf{Part II}: We derive the relation for the spline wavelets $\psi_1(\x;\alpha,\btau)$ and $\psi_5(\x;\alpha,\btau)$ (the rest can be derived identically).  Using proposition \ref{fHT_wavelets}, we immediately arrive at one of the results:  $\H_{\theta_1,\bar \tau} \psi_1(\x;\alpha,\btau)= \H_{0,\bar \tau} \{\psi^{\alpha}_{\tau}(x)\} \beta^{\alpha}_{\tau}(y) = \psi^{\alpha}_{\tau-\bar \tau}(x)\beta^{\alpha}_{\tau}(y)=\psi_1(\x;\alpha,\btau-\bar \tau \boldsymbol u_{\theta_1})$.

The second result relies on the factorization $\H_{\pi/4,\tau}=\H_{0,\tau/2}\H_{\pi/2, \tau/2}$ that holds for functions whose frequency supports are restricted to the quadrants $\{(\omega_x,\omega_y): \omega_x>0, \omega_y>0\}$ and $\{(\omega_x,\omega_y): \omega_x <0, \omega_y<0 \}$. In particular, the condition is satisfied by $\psi_5(\bw; \alpha,\boldsymbol \tau)$ so that, in conjunction with proposition \ref{fHT_wavelets}, we have  
\begin{align*}
\H_{\theta_5,\bar \tau} \psi_5(\x;\alpha,\btau) &=\frac{1}{\sqrt 2} \H_{0,\frac{\bar \tau}{2}}\H_{\frac{\pi}{2},\frac{\bar \tau}{2}} \Big(\psi^{\alpha}_{\tau}(x)\psi^{\alpha}_{\tau}(y)-\psi^{\alpha}_{\tau+\frac{1}{2}}(x)\psi^{\alpha}_{\tau+\frac{1}{2}}(y)\Big) \nonumber \\
&=\frac{1}{\sqrt 2} \Big(\psi^{\alpha}_{\tau-\frac{\bar \tau}{2} }(x)\psi^{\alpha}_{\tau-\frac{\bar \tau}{2}}(y)-\psi^{\alpha}_{\tau-\frac{\bar \tau}{2}+\frac{1}{2}}(x)\psi^{\alpha}_{\tau-\frac{\bar \tau}{2}+\frac{1}{2}}(y)\Big) \nonumber \\
&= \psi_5\Big(\x;\alpha,\btau-\bar{\tau} \frac{1}{\sqrt 2} \boldsymbol u_{\theta_5} \Big). 
\end{align*}

     
\bibliographystyle{amsplain}
\bibliography{bibliography.bib}

\end{document}